# Surface properties of atomically flat poly-crystalline SrTiO$_3$

Sungmin Woo[1], Hoidong Jeong[1], Sang A Lee[1,2], Hosung Seo[3], Morgane Lacotte[4], Adrian David[4], Hyun You Kim[5], Wilfrid Prellier[4], Yunseok Kim[3], Woo Seok Choi[1*]

[1]Department of Physics, Sungkyunkwan University, Suwon, 440-746, Korea

[2]Insitute of Basic Science, Sungkyunkwan University, Suwon, 440-746, Korea

[3]School of Advanced Materials Science and Engineering, Sungkyunkwan University, Suwon, 440-746, Korea

[4]Laboratorie CRISMAT, CNRS UMR 6508, ENSICAEN, Normandie Universite, 6 Bd Marechal Juin, F-14050 Caen Cedex 4, France

[5]Department of Nanomaterials Engineering, Chungnam National University, Daejeon, 305-764, Korea

*e-mail: choiws@skku.edu.

Comparison between single- and the poly-crystalline structures provides essential information on the role of long-range translational symmetry and grain boundaries. In particular, by comparing single- and poly-crystalline transition metal oxides (TMOs), one can study intriguing physical phenomena such as electronic and ionic conduction at the grain boundaries, phonon propagation, and various domain properties. In order to make an accurate comparison, however, both single- and poly-crystalline samples should have the same quality, e.g., stoichiometry, crystallinity, thickness, etc. Here, by studying the surface properties of atomically flat poly-crystalline SrTiO$_3$ (STO), we propose an approach to simultaneously fabricate both single- and poly-crystalline epitaxial TMO thin films on STO substrates. In order to grow TMOs epitaxially with atomic precision, an atomically flat, single-terminated surface of the substrate is a prerequisite. We first examined (100), (110), and (111) oriented single-crystalline STO surfaces, which required different annealing conditions to achieve atomically flat surfaces, depending on the surface energy. A poly-crystalline STO surface was then prepared at the optimum condition for which all the domains with different crystallographic orientations could be successfully flattened. Based on our atomically flat poly-crystalline STO substrates, we envision expansion of the studies regarding the TMO domains and grain boundaries.

## Introduction

Physical properties of crystals are mostly determined by the periodic ordering of atoms, *i.e.*, the translational symmetry. According to their atomic structures, materials can be classified into three categories, *i.e.*, amorphous, poly-, and single-crystalline. Although comprised of the same elements, many characteristic of a material, such as mechanical, optical, thermal, electric, magnetic and chemical properties, can be distinguished among the different structures.

A comparison between the different atomic structures provides important clues to understanding the nature of the materials. For example, by comparing the single- and poly-crystalline structures, the precise role of the structural grain boundaries can be investigated. At the boundaries, the electron and phonon conductions are hampered due to the breaking of the translational symmetry. If one can resolve the role of the structural boundaries in scattering electrons and phonons separately, a novel phonon-glass-electron crystal might be discovered as an efficient thermoelectric material, which preferentially scatters the phonon at the boundaries.[1] On the other hand, the role of the crystal domains in the poly- or bi-crystals can also be investigated by comparison with single-crystals, which might also lead to novel device performance.[2-5] For example, magnetic domains in the poly-crystal perovskite manganites $La_{1-x}AE_xMnO_{3-\delta}$ ($AE$ = Ca, Sr, or vacancies) mostly switch independently in a field, greatly influencing the magnetoresistance (MR) property.[2,6,7]

Practically, however, it is rather challenging to compare the single- and poly-crystalline samples on an equal footing, since the growth conditions of these different structures are usually different. In general, the optimum growth conditions for high-quality single-crystalline samples do not result in the formation of poly-crystalline samples with similar quality, *i.e.*, high quality poly-crystals form at a different condition. Therefore, while high quality single-crystals are often accessible, poly-crystals with grains of the optimum crystalline quality are rare. For complex materials such as ternary transition metal oxides (TMO), the situation is worse, as the stoichiometry strongly depends on the growth conditions.[8]

In order to overcome such difficulties, we propose using atomically smooth poly-crystalline TMOs as the substrate for epitaxial poly-crystalline TMO oxide growth. The poly-crystalline substrates enable the growth of high-quality poly-crystalline thin films at optimum conditions, based on the epitaxial relationship between the substrate and the film. It should be noted that a similar approach has been adopted using polycrystalline substrates such as $TiO_2$, $LaAlO_3$, and $Sr_2TiO_4$.[9-12] The original purpose of this so-called "combinatorial substrate epitaxy (CSE)" is to determine the phase and orientation relationships between the substrate and the deposited film for all possible orientations in a single sample, and to screen a particular physical property as a function of orientation. On the other hand, studies involving a direct comparison between single- and poly-crystalline thin films using this method have been rarer.[2,3] In addition, in order to investigate epitaxial TMO thin films and heterostructures with atomic precision, such as artificial superlattices and functional interfaces, an atomically smooth surface of the substrate is necessary.[13,14] Atomically flat surfaces enable epitaxial layer-by-layer growth in the atomic scale, which results in homogeneously terminated surfaces. While thin films and heterostructures fabricated in this way should provide a novel route to investigate the physics of the grains and the grain boundaries, there has not been any effort to characterize the atomic structure of the polycrystalline oxide surfaces.

In this paper, we investigate the surface properties of poly-crystalline $SrTiO_3$ (STO) and pursue an atomically smooth surface for poly-crystalline TMO thin film growth. STO was chosen as it is the most commonly used substrate for the growth of perovskite-type TMO thin films and heterostructures. The poly-crystalline perovskite substrate is comprised of many different crystallographic orientations with different surface properties, and therefore, treatment conditions for each crystalline orientation should be known to obtain an atomically smooth poly-crystalline surface. Since it was impossible to test the surface treatment conditions for all the different orientations individually, we first investigated three representative crystallographic orientations, *i.e.*, (100), (110), and (111), of single-crystalline substrates. These three orientations have distinct topmost atomic layers with different surface energy,[15] where the surfaces of poly-crystalline STO can be considered as a combination of the different atomic layers. Indeed, we achieved an atomically smooth poly-crystalline STO surface by using the optimum surface treatment condition. The realization of atomically smooth poly-crystalline STO surface was discussed in the context of the crystallographic orientation, surface energy (potential), and topography of the grains.

Results and discussion

Surface properties of single-crystalline STO. Previously, there have been many studies on the preparation of atomically flat single-crystalline STO surfaces using chemical etching and subsequent annealing.[15-21] However, a systematic study on the surface treatment for the representative STO single crystalline surfaces, *i.e.*, (100), (110), and (111) surfaces has been rare. Moreover, the surface treatment of poly-crystalline STO has never been investigated, to the best of our knowledge.[22]

Figure 1 shows (001) surfaces of STO single crystal under different surface treatment conditions. By varying the annealing temperature, annealing time, and etching time, we could obtain an optimal surface treatment window to obtain an atomically flat surface. The atomically smooth surface could be depicted from the typical step-and-terrace structure of the $TiO_2$-terminated STO surface with an average height of ~3.9 Å, from the atomic force microscopy (AFM) analysis. The step-and-terrace structure originates from the miscut angle when the crystal surfaces are prepared. Each condition, *i.e.*, annealing temperature, annealing time, and etching time, has a different role in modifying the surface quality. For example, Fig. 1(a) shows annealing temperature dependence, which affects the roughness of the STO surface. As the annealing temperature increases from 900 to 1000 to 1200°C, the RMS roughness decreases from 0.123 to 0.078 to 0.051 nm. Lippmaa *et al.* suggest that a higher annealing temperature straightens the step edge and reduces the island structures.[23] In the 900-1200°C range, we indeed observed a systematic decrease in the island structures along with a reduced surface roughness.

On the other hand, a longer annealing time is known to result in SrO enrichment on the surface, which degrades the single termination.[17, 24] However, in the range studied here, *i.e.* four to eight hours, SrO surface enrichment could not be observed, as clearly shown in Fig. 1(b). Instead, we observed a straightening of the step edges along with reduced RMS roughness values, possibly due to the self-organization of the atoms on the surface with longer annealing times.[23] Finally, we confirmed that the moderate etching time (15 to 420 seconds) does not affect the step-and-terrace structure of the surface as shown in Fig. 1(c). As reported previously, chemical etching only dissolves the SrO layer on the STO surface.[16, 25]

Similar surface treatment experiments were repeated for the three representative crystallographic orientations, *i.e.*, (100), (110), and (111) surfaces, of STO. First, we note that the surfaces with different orientations have distinct surface energies, which results in different surface treatment conditions. Table 1 presents the calculated surface energy of the (100), (110), and (111) facets with different atomic terminations. Note the significant differences in the surface energy between the studied crystallographic orientations, while the role of the atomic termination is relatively marginal. If we compare among the Ti-containing top most layers, the (100), (110), and (111) surfaces have a surface energy of 6.840, 20.016, and 28.596 eV/nm$^2$. As anticipated, the non-polar (100) facet has the lowest surface energy, compared to the two other polar surfaces, the (110) and (111) facets, suggesting that it is energetically most favored and an atomically flat surface would be easily obtainable.[26] We also note a substantial difference in the surface energy between the (110) and (111) surfaces (more than 40%). The energy difference indicates that the annealing conditions for obtaining atomically flat surfaces should be different for the (110) and (111) STO surfaces.

The result of substrate treatment experiment for the single-crystalline STO surfaces for (100), (110), and (111) surfaces are summarized in Fig. 2. Each orientation has its own window for forming an atomically flat step-and-terrace structure. Figure 2(a) shows the result of surface condition as a function of annealing time and temperature. We neglect the etching time dependence, as it does not affect the surface topology as shown in Fig. 1(c). The filled symbols in Fig. 2(a) indicate atomically flat surfaces, while the unfilled symbols indicate rough surfaces. Red, yellow, and blue symbols indicate (100), (110), and (111) oriented surfaces, respectively. The respectively colored windows show the range where the atomically flat surfaces have been observed for each surface orientation. As anticipated from Table 1, the (100) STO surface has the largest area, owing to its smallest surface energy. The (110) and (111) surfaces have somewhat smaller area, suggesting that they require higher temperature or longer annealing time to obtain atomically flat surfaces. Beneath the filled symbols, we also present the RMS roughness values of each terrace in nm, for the atomically flat surfaces. The general trend of the RMS roughness indicates that smoother surfaces can be obtained for higher annealing temperatures and longer annealing times, regardless of the surface orientations. This is consistent with what we have observed in Fig. 1 for the (100) surface.

The representative AFM topographic images of (100), (110), and (111) single-crystalline STO surfaces etched for 15 seconds and subsequently annealed at 1000°C for 6 hours (This condition is indicated as black dashed circles in Fig. 2(a).) are shown in Fig. 2(b), 2(c), and 2(d). At this condition, atomically flat step-and-terrace structures can be observed for the differently orientated STO surfaces. The line profiles of (100), (110), and (111) surface show an average step height of 0.381, 0.275, and 0.213 nm, respectively, as shown in the plots below each image. These values correspond to the theoretical step height values of 0.391, 0.276, and 0.225 nm, respectively for (100), (110), and (111) surfaces.[15] The correspondence suggests that the observed surfaces are all single-terminated either by $TiO_2$ or SrO for (100), $SrTiO^{4+}$ or $O_2^{4-}$ for (110), and $Ti^{4+}$ or $SrO_3^{4-}$ for the (111) surface as graphically demonstrated in Table 1. In addition, due to the use of buffered HF chemical etching which selectively resolves the Sr-containing layer (or Sr compound), the surfaces are likely to be Ti-rich or Ti terminated.[15,25] We also note that a polar surfaces might induce different crystalline phases or attract more adsorbents with time, compared to a non-polar surfaces.[27] However, we did not observe such aging process within the time frame of our experiment (several weeks).

**Atomically flat poly-crystalline STO surface.** Based on the experiments regarding single-crystalline substrate, we extended our research on poly-crystalline STO surfaces. Figure 3 shows a high-resolution x-ray diffraction (XRD) $\theta$-$2\theta$ scan of the polycrystalline STO, indicating a good crystalline quality with grains of various crystallographic orientations. We could observe all the polycrystalline STO peaks located at low angle up to $2\theta = 90°$, which were expected from the various crystallographically orientated grains.

Using the condition used to treat single-crystalline STO with the representative surface orientations (the condition indicated as dashed circle in Fig. 2(a)), we obtained poly-crystalline STO with atomically flat surfaces. While there is a possibility that domains with high miscut angles might need even higher energies to form an atomically smooth surface, we confirmed that the conditions obtained in Fig. 2 should be sufficient for most of the grains.

Figure 4 shows the electron backscattering diffraction (EBSD) color map, AFM topography, and kelvin probe force microscopy (KPFM) potential images of chemically-treated and annealed poly-crystalline STO surface. The different surface measurements were performed at the same corresponding spot. The EBSD color map (Fig. 4(a)) indicates that each domain indeed has different crystallographic orientations. The inverse pole figure on the right of Fig. 4(a) provides orientation information regarding domains indicated as black squares in the EBSD map. (The exact (*hkl*) values of the surfaces can be found in the Supplementary Information.) These domains have been selected randomly while measuring the AFM. Since preparation of EBSD measurement accompanies serious surface contamination in the atomic scale which prevented subsequent AFM measurements, we measured the AFM topography first for the randomly selected domains, and then confirmed the surface orientations of the domains using EBSD. AFM topographic and KPFM potential images in Figs. 4(b) and 4(c), respectively, show the corresponding shapes of the domains. We observed distinct changes between some of the domains in the KPFM image, indicating surface potential changes between the differently oriented domains.

In order to study the relationship between the surface property and crystallographic orientation, we measured the detailed surface topology for some of the selected domains. Figures 5(a)-5(f) show magnified AFM images for the selected domains indicated as black squares in Fig. 4. The surfaces of the STO could be categorized into two different groups with atomically smooth surfaces. First, Figs. 5(a)-(d) show surfaces with step-and-terrace structures. The line profile of the surface with step-and-terrace structure (Fig. 5(g)) indicates that the typical height of the steps (~1 nm) is larger than that of the single crystalline STO surfaces (0.2 – 0.4 nm), possibly due to the large miscut-induced step bunching (Fig. 5(g)). Nevertheless, judging from the flatness of each terrace, the surfaces can be considered as atomically flat. Second, Figs. 5(e) and 5(f) show surfaces with extremely low surface roughness (less than a unit cell), as shown in the line profile in Fig. 5(h). The first group (Figs. 5(a), 5(b), 5(c), and 5(d)) has an RMS roughness of 6.524, 7.499, 5.797, and 9.761 nm, (Here, the RMS roughness is for the whole image (0.5 × 0.5 $\mu m^2$), as the width of individual terraces were too small.) with surface potential energy of -142.383, -145.854, -146.597, and -153.638 mV, respectively. On the other hand, the second group (Figs. 5(e) and 6(f)) has much lower RMS roughness of 0.299 and 0.234 nm, (for the whole image) with lower surface potential of -166.851 and -172.306 mV, respectively.

Based on the inverse pole figure shown in Fig. 4(a), we conclude that the first group (surfaces with the step-and-terrace structure) is located farther away than the second group (surfaces without the step-and-terrace structure) from the (100) cubic surface, which has the lowest surface energy (Table 1). The increased surface potential (Fig. 4(c)) and energy in the first group seems to induce step bunching and the resultant step-and-terrace structure. It was rather unexpected that the second group did not show the one unit cell step-and-terrace structure observed for a single-crystalline (100) STO surface, particularly since they have substantially low surface RMS roughness. We note, however, that even those surfaces have a modest miscut angle compared to the single-crystalline (100) surface, and therefore, the surface quality cannot be the same.

## Summary

We studied various surface properties of atomically flat poly-crystalline STO. First, we obtained a single optimum surface treatment condition for the single-crystalline (100), (110), and (111) STO substrates. This condition was used to obtain atomically flat poly-crystalline STO surfaces. The EBSD, AFM, and KPFM analyses suggest that the domains with surface orientations close to (100) cubic have a low surface energy, which induces extremely flat surfaces, while the domains with surface orientations away from (100) cubic have higher surface energy, which induces substantial step bunching of the surfaces. Our observations concerning the formation of atomically flat surfaces of poly-crystalline STO provide an avenue for studying epitaxial poly-crystalline perovskite TMO thin films with intriguing physical properties.

## Methods

**Sample fabrication.** Poly-crystalline STO was made of equimolar amounts of $SrCO_3$ and $TiO_2$ powders ($SrCO_3$, Aldrich and $TiO_2$, Cerac with 99.9% purity). The powders were weighted in stoichiometric proportions, mixed intimately, and reacted in their solid states using thermal treatments.[11,28] The precursors were annealed at 1200°C for 14 h, to obtain the desired perovskite phase. An additional step of grinding was necessary to obtain powders with homogeneous grain sizes. The calcined powders were loaded into a 20 mm graphite die, and graphite paper was added around and on the top of the punches to protect the powders from external pollution. The resulting STO powders were sintered under 16 MPa at 1400°C for 20 min using a spark plasma sintering (SPS) apparatus (Struers Tegra Force-5). The heating (cooling) rate was 100°C per minute, with a simultaneous increase (decrease) of the uniaxial load. The STO poly-crystals were cut from the SPS ceramic and mechanically polished by Shinkosha Co. Ltd., resulting in a mirror-like surface with an RMS roughness of 1.521 nm.

**Surface preparation.** In order to obtain well-defined, atomically flat surfaces, we carried out chemical etching and thermal annealing for the single- and poly-crystalline STO substrates. First, the substrates were cleaned by soaking in deionized-water for 10 seconds. Then the surfaces were etched in a buffered hydrogen fluoride (HF) $NH_4F:HF = 10:1$ solution for 15-420 seconds. The remaining chemical fluid was removed by spraying deionized-water on the substrates. After chemical etching, the substrates were annealed at 900~1200°C in air for 2-8 hours to pursue a smooth step-and-terrace structure.

**Surface property measurements.** AFM (Park Systems NX10) with a Si probe tip (Budget sensors ContAl-G) was used to examine the surface topography.[29] KPFM was used for measuring surface potential. To determine the surface potential, ac modulation and dc voltages were applied to the tip or the sample. When the ac modulation voltage is applied to the tip in non-contact mode, the tip oscillates by the force between the tip and the sample surface. This force is composed of the van der Walls force, the electrostatic force, and the force, which vibrates the tip. The surface potential $V_s$ is obtained from the electrostatic force $F_{es}$ as follows:

$$F_{es} = -\frac{1}{2}(\frac{\partial C}{\partial z})\left\{(V_s+V_{dc})^2 + \frac{V_{ac}^2}{2} + 2(V_s+V_{dc})\sin(\omega t) + \frac{V_{ac}^2}{2}\cos(2\omega t)\right\}$$

If the dc voltage $V_{dc}$ is equal to the surface potential $V_s$, $\omega$ component of the electrostatic force is vanished. This dc voltage at the nullifying of the $\omega$ component is shown as a surface potential. For the KPFM analysis, an Au coated tip (Nanosensors PPP-NCHAu) was used. To confirm the structural quality of the poly-crystalline STO, we performed XRD, Rigaku SmartLab and EBSD, JSM7000F. For the EBSD, the sample was mounted at a 70° angle from scanning electron microscope operated at 20 kV. The probe current of the aperture was $1\times10^{-8}$ A and the Kikuchi diagrams were recorded with the beam step size of 0.5 $\mu$m. For image processing, a comercial program (OIM Analysis 5.31) was used.

**Theoretical calculation.** For the surface energy calculation for (100), (110), and (111) facets of STO, we performed spin-polarized density functional theory (DFT) calculations with the VASP code[30] and the PBE[31] exchange-correlation functional. Detailed description of the computational methods and surface modeling can be found in the Supplementary Information.

## Acknowledgements

We appreciate valuable discussions with Jaeyeol Hwang, Paul A. Salvador and Yeong Jae Shin. This work was supported by the Basic Science Research Program through the National Research Foundation of Korea (NRF) funded by the Ministry of Science, ICT and future Planning (NRF-2014R1A2A2A01006478) and by the Ministry of Education (NRF-2013R1A1A2057523). YK was supported by Faculty Research Fund, Sungkyunkwan University 2012. The theoretical calculation was carried out at the Center for Functional Nanomaterials, Brookhaven National Laboratory, which is supported by the U.S. Department of Energy, Office of Basic Energy Sciences, under Contract No. DE-AC02-98CH10886.

## Author contributions

S.W, H. J, and S.A.L conducted the experiment and analyzed the results. S.W, H.S and Y.K measured the surface property using AFM and KPFM. M.L, A.D, and W.P fabricated the poly-crystalline STO samples. H.Y.K performed the DFT calculation. S.W and W.S.C wrote the main manuscript and all authors reviewed the manuscript. W.S.C initiated and supervised the research.

Figure legends

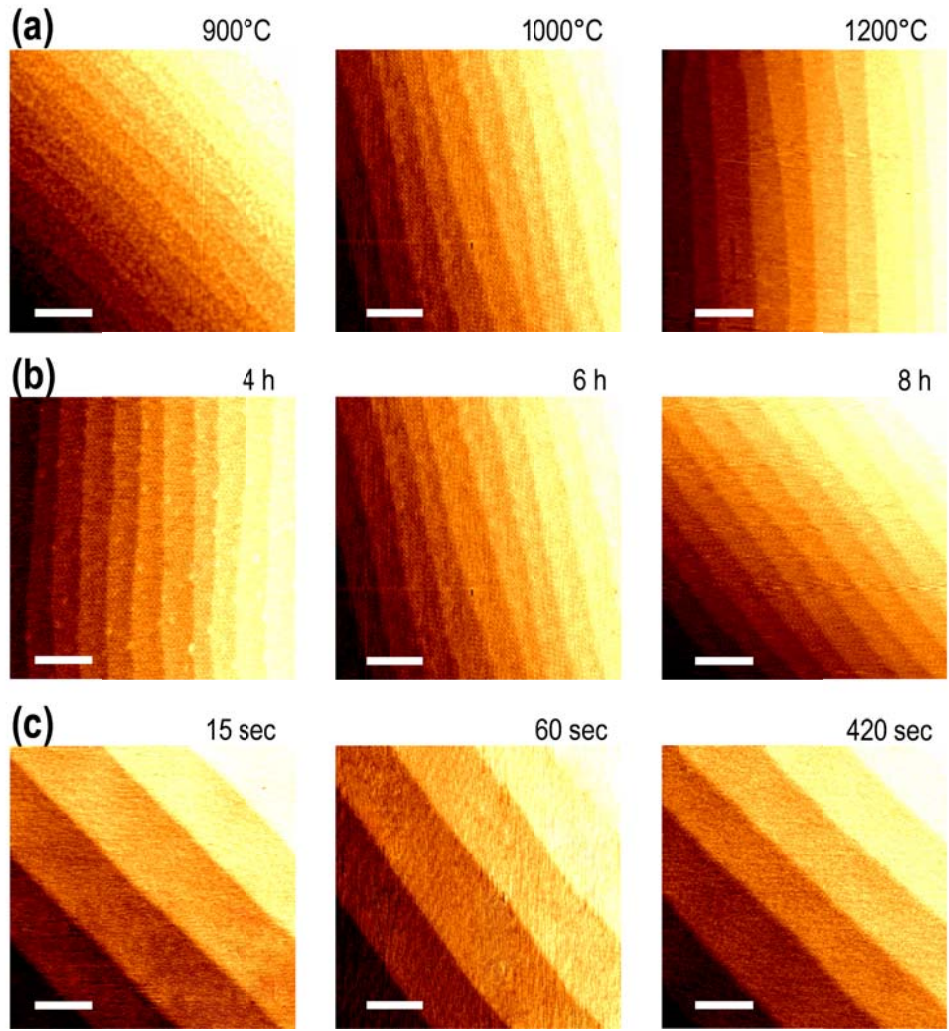

Figure 1 | AFM topographic images for chemically etched and subsequently annealed STO (001) surfaces. (a) STO (100) surfaces etched for 15 seconds and annealed for 6 hours at different temperatures. Samples annealed at 900 (RMS roughness of the terrace = 0.123 nm), 1000 (0.078 nm), and 1200°C (0.051 nm) indicate that they become smoother by increasing the annealing temperature. (b) STO (100) surfaces etched for 15 seconds and annealed at 1000°C for 4 (RMS roughness of the terrace = 0.115 nm), 6 (0.098 nm), and 8 (0.077 nm) hours. The step edges became straight and the RMS roughness of the terrace decreased as the annealing time increases. (c) STO (100) surfaces etched for different amounts of time and subsequently annealed at 1000°C for 6 hours. Etching for 15

(RMS roughness of the terrace = 0.063 nm), 60 (0.076 nm), and 420 (0.067 nm) seconds did not significantly change the surface quality. The scale bars indicate 0.2 μm.

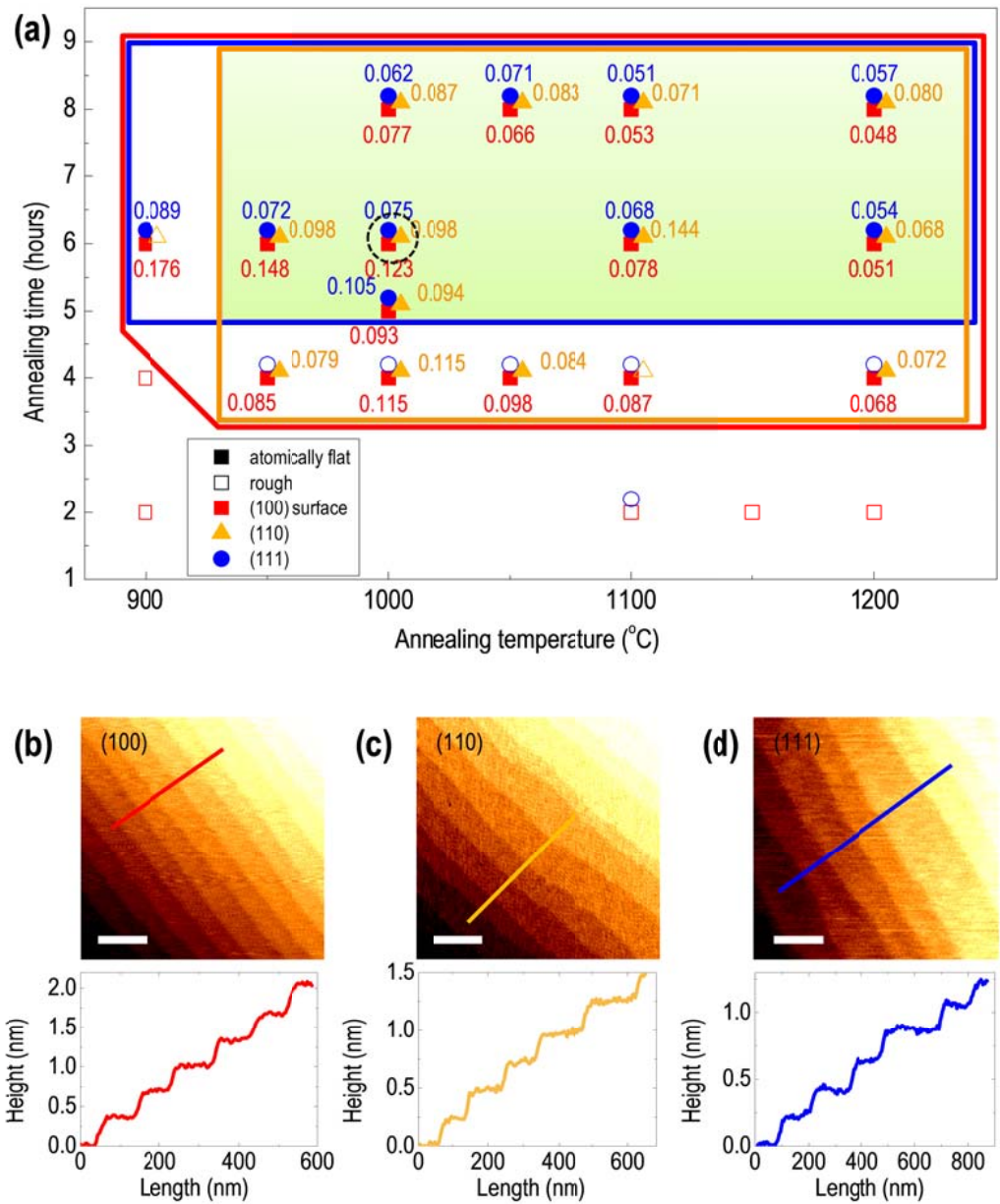

Figure 2 | Summary of obtaining atomically flat STO surfaces for different crystallographic orientations. (a) Result of annealing as a function of annealing time and temperatures. Filled (empty) symbols indicate that atomically flat surfaces have (not) been obtained using the designated annealing condition. Red squares, yellow triangles, and blue circles indicate (100), (110), and (111) STO surfaces, respectively. The red, yellow, and blue regions indicate where the atomically flat surfaces

with step-and-terrace structures have been obtained for STO. The numbers next to the symbols indicate the RMS roughness values of the terrace in nm. AFM topographic images for (b) (100), (c) (110), and (d) (111) surfaces with line profiles. The scale bars indicate 0.2 $\mu$m. These samples have been etched for 15 seconds, and subsequently annealed at 1000 °C for 6 hours (indicated by the dashed circle in (a)).

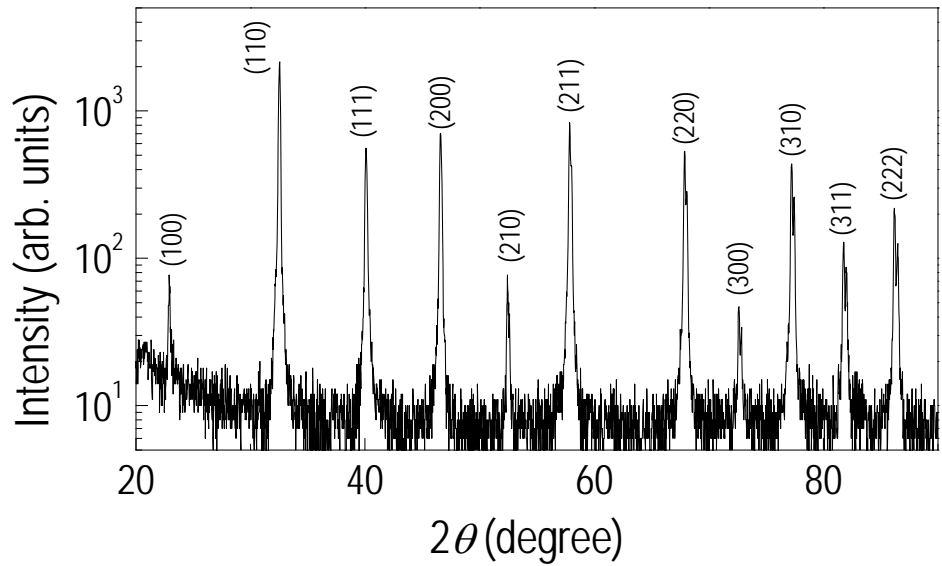

Figure 3 | Structural quality of poly-crystalline STO. High-resolution XRD $\theta$-$2\theta$ scan for poly-crystalline STO showing Bragg peaks expected for different crystallographic orientations.

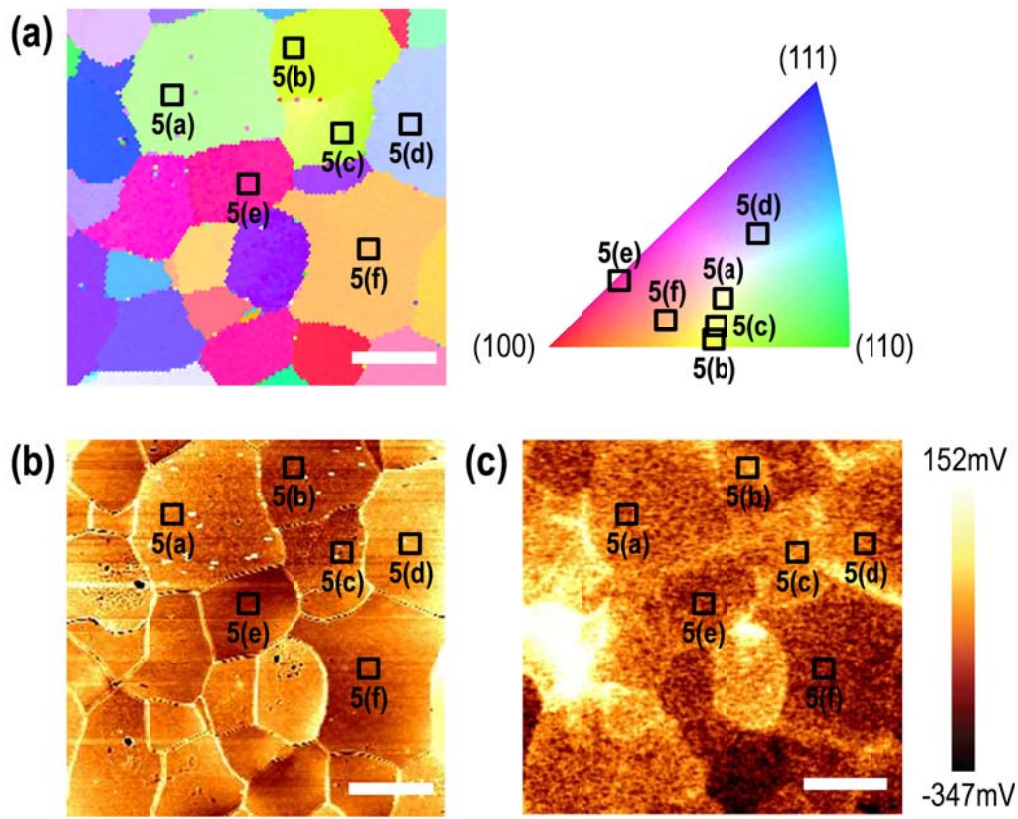

Figure 4 | Corresponding domain images of the poly-crystalline STO surface obtained by (a) EBSD, (b) AFM, and (c) KPFM. Inverse pole figure color map in (a) indicates each domain for which we measured the high resolution AFM shown in Fig. 5. The scale bars represent 10 $\mu$m.

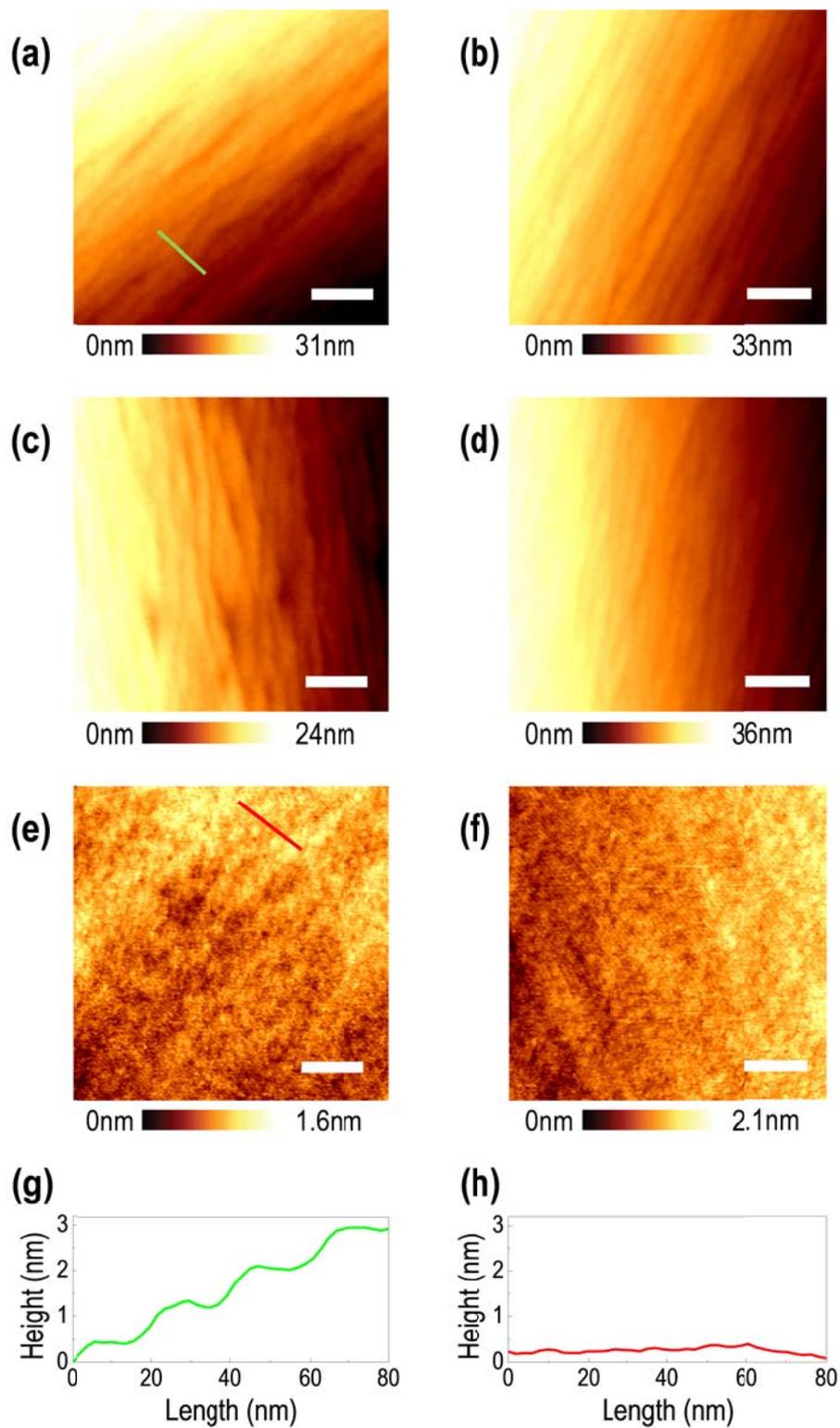

Figure 5 | AFM topographic images of poly-crystalline STO surface. The sample was treated using the condition indicated as dashed circle in Fig. 2(a) (15 seconds etching and 6 hours annealing at 1000°C.) All the domains with different crystallographic orientations have atomically smooth surfaces, and are

from the spots shown in Fig. 4. (a), (b), (c), and (d) show step-and-terrace structure with an average step height of a few nanometers, indicating step bunching for highly miscut perovskite surfaces. (e) and (f) show low RMS roughness, indicating that the surfaces are atomically flat. The scale bars indicate 0.1 $\mu$m. (g) and (h) show line profiles of (a) and (e) respectively, indicating atomically smooth STO surfaces.

Table 1 | DFT-calculated surface energy of STO surfaces with different crystallographic orientations and surface terminations. The surface energy significantly depends on the crystallographic orientation, while the surface termination has a smaller effect. The STO (100) facet is energetically the most favored, confirming our experimental results.

| Surface orientation | (100) | | (110) | | (111) | |
|---|---|---|---|---|---|---|
| Termination | $TiO_2$ | SrO | SrTiO | $O_2$ | Ti | $SrO_3$ |
| Surface energy (eV/nm$^2$) | 6.840 | 6.858 | 20.016 | 20.467 | 28.596 | 34.078 |